\documentclass[12pt]{article}
\usepackage[left=2.5cm,top=2.50cm,right=2.5cm,bottom=2.50cm]{geometry}
\usepackage{mathrsfs}
\usepackage{amsmath,amssymb,latexsym,color,cancel,graphicx,bbm,colortbl}
\usepackage[english]{babel}
\usepackage[latin1]{inputenc}
\usepackage{ragged2e}
\usepackage{cite}
\begin{document}
\date{}

\title{An algebraic approach to a charged particle in a uniform magnetic field}
\author{D. Ojeda-Guill\'en$^{a}$,\footnote{{\it E-mail address:} dojedag@ipn.mx} \ M. Salazar-Ramírez$^{a}$,\\ R. D. Mota$^{b}$ and V. D. Granados$^{c}$} \maketitle

\begin{minipage}{0.9\textwidth}
\small $^{a}$ Escuela Superior de C\'omputo, Instituto Polit\'ecnico Nacional,
Av. Juan de Dios B\'atiz esq. Av. Miguel Oth\'on de Mendiz\'abal, Col. Lindavista,
C.P. 07738, Ciudad de M\'exico, Mexico.\\

\small $^{b}$ Escuela Superior de Ingenier{\'i}a Mec\'anica y El\'ectrica, Unidad Culhuac\'an,
Instituto Polit\'ecnico Nacional, Av. Santa Ana No. 1000, Col. San
Francisco Culhuac\'an, C.P. 04430, Ciudad de M\'exico, Mexico.\\

\small $^{c}$ Escuela Superior de F{\'i}sica y Matem\'aticas, Instituto Polit\'ecnico Nacional,
Ed. 9, Unidad Profesional Adolfo L\'opez Mateos, Col. Zacatenco,
C.P. 07738, Ciudad de M\'exico, Mexico.\\

\end{minipage}

\begin{abstract}
We study the problem of a charged particle in a uniform magnetic field with two different gauges,
known as Landau and symmetric gauges. By using a similarity transformation in terms of the displacement operator
we show that, for the Landau gauge, the eigenfunctions for this problem are the harmonic oscillator number coherent states.
In the symmetric gauge, we calculate the $SU(1,1)$ Perelomov number coherent states for this problem
 in cylindrical coordinates in a closed form. Finally, we show that these Perelomov number coherent states
are related to the harmonic oscillator number coherent states by the contraction of the $SU(1,1)$ group to the Heisenberg-Weyl group.

\end{abstract}


\section{Introduction}

Harmonic oscillators coherent states were introduced by Schr\"odinger at the beginning of the quantum mechanics \cite{Schrocoherent}. Glauber defined
these states as the eigenfunctions of the annihilation operator \cite{glauber}. Klauder showed that these states are obtained by applying the Weyl operator
to the harmonic oscillator ground state \cite{Klau}. Harmonic oscillator coherent states are gaussian functions displaced from origin which maintain their
shape over time. Boiteux and Levelut defined the number coherent states for the harmonic oscillator by applying the Weyl operator to any excited state \cite{BandL}. These states are called displaced number states or number coherent states and were extensively studied in the middle of the last century. Most of their properties are compiled in references \cite{sen,ple,hus,eps}. In reference \cite{Nieto2}, Nieto review these states and gave their most general form.

Perelomov generalized the Klauder coherent states to any Lie group by applying the group displacement operator to the
lowest normalized state \cite{Perel}. The Perelomov coherent states haven been applied to many physical problems as can be seen in references \cite{Perellibro,Klauderlibro,Gazeaulibro}. Gerry defined the $SU(1,1)$ number coherent states by applying the Perelomov displacement operator to any excited state and used this definition to calculate the Berry's phase in the degenerate parametric amplifier \cite{gerryberry}. Recently, we have studied the Perelomov number coherent states for the $SU(1,1)$ and $SU(2)$ groups. In particular, we gave the most general expression of these states, their ladder operators and applied them to calculate the eigenfunctions of the non-degenerate parametric amplifier \cite{nosotros1} and the problem of two coupled oscillators \cite{nosotros2}. In reference \cite{nosotros3}, we computed the number radial coherent states for the generalized MICZ-Kepler problem by using
the $su(1,1)$ theory of unitary representations and the tilting transformation.

On the other hand, the problem of a charged particle in a uniform magnetic field has been widely studied in classical mechanics, condensed matter physics, quantum optics and relativistic quantum mechanics, among others. The energy spectrum of this problem is known as the Landau levels. The interaction of an electron with the uniform magnetic field is described by means of electromagnetical potentials. However, different gauges give raise to the same electromagnetic field \cite{Capri}. The coherent states for this problem have been obtained previously by using different formalisms, as can be seen in references \cite{Malkin,Feldman,Fakhri,Fakhri2,Kowals,Dehghani}.

The aim of this work is to introduce an algebraic approach to study the problem of a charged particle in an uniform magnetic field and obtain its coherent states. Specifically, the algebraic approach used in this
work is the tilting transformation, which offers to graduate students an alternative method to the commonly used analytical approach, in order to study and exactly solve several problems in quantum mechanics.

This work is organized as it follows. In Section $2$, we give a summary on the Heisenberg-Weyl and $SU(1,1)$ groups and
its number coherent states. In Section $3$, we study the problem of a charged particle in a uniform magnetic field in cartesian coordinates with the
Landau gauge. We solve this problem and show that its eigenfunctions are the harmonic oscillator number coherent states. In Section $4$, we study the Landau levels problem in cylindrical coordinates with the symmetric gauge. We construct the $SU(1,1)$ Perelomov number coherent states for its eigenfunctions. In Section $5$, we contract the $SU(1,1)$ group to the Heisenberg-Weyl group. We show that, under this contraction the $SU(1,1)$ Perelomov coherent states are related to the harmonic oscillator coherent states. Finally, we give some concluding remarks.

\section{$H(4)$ and $SU(1,1)$ number coherent states}

\subsection{Heisenberg-Weyl group}

The annihilation and creation operators of the harmonic oscillator $a, a^{\dag}$, together with the number and identity operators $a^{\dag}a,I$ satisfy the following relations
\begin{equation}
[a,a^{\dag}]=I,\quad\quad [a,a^{\dag}a]=a,\quad\quad [a^{\dag},a^{\dag}a]=-a^{\dag},\quad\quad [a,I]=[a^{\dag}a,I]=0=[a^{\dag},I].
\end{equation}
These equations are known as the Heisenberg-Weyl algebra $h(4)$. The action of these operators on the Fock states is given by
\begin{equation}
a^{\dag}|n\rangle=\sqrt{n+1}|n+1\rangle, \quad\quad a|n\rangle=\sqrt{n}|n-1\rangle, \quad\quad a^{\dag}a|n\rangle=n|n\rangle.
\end{equation}
The harmonic oscillator coherent states are defined in terms of these operators as
\begin{equation}
|\alpha\rangle=D(\alpha)|0\rangle=e^{\alpha a^{\dag}-\alpha^*a}|0\rangle=\exp{\left[-\frac{1}{2}|\alpha|^2\right]}\sum_0^\infty\frac{\alpha^n}{\sqrt{n!}}|n\rangle,
\end{equation}
where $D(\alpha)$ is the Weyl operator (also called displacement operator), $|0\rangle$ is the ground state and $\alpha$ is a complex number given by
\begin{equation}
\alpha=\sqrt{\frac{m\omega}{\hbar}}x_0+\frac{i}{\sqrt{2m\omega\hbar}}p_0.
\end{equation}
This unitary operator $D(\alpha)$ can be expressed in a disentangled form by using the Weyl identity as follows \cite{Davydov}
\begin{equation}
D(\alpha)=e^{-|\alpha|^2/2}e^{\alpha a^{\dag}}e^{-\alpha^*a}.\label{disentangled}
\end{equation}
The harmonic oscillator number coherent states are defined as the action of the Weyl operator on any excited state $|n\rangle$ \cite{BandL}
\begin{equation}
|n,\alpha\rangle=D(\alpha)|n\rangle=e^{-|\alpha|^2/2}e^{\alpha a^{\dag}}e^{-\alpha^*a}|n\rangle.
\end{equation}
By using the Baker-Campbell-Hausdorff identity
\begin{equation}
e^{-A}Be^A=B+\frac{1}{1!}[B,A]+\frac{1}{2!}[[B,A],A]+\frac{1}{3!}[[[B,A],A],A]+...,\label{BCH}
\end{equation}
and the commutation relationship of the ladder operators $[a,a^{\dag}]=1$, it can be shown the following properties
\begin{eqnarray}
A(\alpha)=D^{\dag}(\alpha)a D(\alpha)=a+\alpha,\label{A}\quad\quad
A^{\dag}(\alpha)=D^{\dag}(\alpha)a^{\dag}D(\alpha)=a^{\dag}+\alpha^*.\label{Ad}
\end{eqnarray}
These operators play the role of annihilation and creation operators when they act on
the harmonic oscillator number coherent states, since \cite{BandL}
\begin{eqnarray}
A^{\dag}|n,\alpha\rangle=\sqrt{n+1}|n+1,\alpha\rangle,\quad\quad A|n,\alpha\rangle=\sqrt{n}|n-1,\alpha\rangle.
\end{eqnarray}
By using the disentangled form of the Weyl operator $D(\alpha)$ of equation (\ref{disentangled}), we can prove that the most general
form of these states in the Fock space is \cite{Nieto2}
\begin{equation}
|n,\alpha\rangle=e^{-|\alpha|^2/2}\sum_{k=0}^{\infty}\frac{(\alpha a^{\dag})^k}{k!}\sum_{j=0}^n\frac{(-\alpha^*a)^j}{j!}
\left(\frac{(n-j+k)!n!}{(n-j)!(n-j)!}\right)^{1/2}|n-j+k\rangle.\label{nieto}
\end{equation}

On the other hand, the eigenfunctions of the one-dimensional harmonic oscillator are given by
\begin{equation}
\psi_n(x)=N_nH_n(\beta x)e^{-\frac{1}{2}\beta^2x^2},\label{harmonic}
\end{equation}
where $H_n(\beta x)$ are the Hermite polynomials and $\beta=\sqrt{\frac{m\omega}{\hbar}}$, $N_n=\left(\frac{\beta}{\pi^{1/4}2^nn!}\right)^{1/2}.$
The Weyl operator $D(\alpha)$ can be expressed in terms of the harmonic oscillator position $x$ and momentum $p$ operators as
\begin{equation}
D(x_0,p_{0x})=e^{\frac{i}{\hbar}(p_{0x}x-x_0p_x)}=e^{-\frac{ix_0p_{0x}}{2\hbar}}e^{\frac{ip_{0x}x}{\hbar}}e^{\frac{ix_0p_x}{\hbar}},
\end{equation}
as it is shown in reference \cite{Davydov}. Thus, the action of this operator on the harmonic oscillator eigenfunctions of equation (\ref{harmonic}) is
\begin{equation}
D(x_0,0)\psi_n(x)=N_ne^{-\frac{1}{2}\beta^2(x-x_0)^2}H_n\left(\beta(x-x_0)\right).\label{dfunction}
\end{equation}

\subsection{$SU(1,1)$ group}

The $su(1,1)$ Lie algebra is generated by the set of operators $\{K_{+}$, $K_{-}, K_{0}\}$. These operators satisfy
the commutation relations \cite{Vourdas}
\begin{eqnarray}
[K_{0},K_{\pm}]=\pm K_{\pm},\quad\quad [K_{-},K_{+}]=2K_{0}.\label{com}
\end{eqnarray}
The action of these operators on the Fock space states $\{|k,n\rangle, n=0,1,2,...\}$ is given by
\begin{equation}
K_{+}|k,n\rangle=\sqrt{(n+1)(2k+n)}|k,n+1\rangle,\label{k+n}
\end{equation}
\begin{equation}
K_{-}|k,n\rangle=\sqrt{n(2k+n-1)}|k,n-1\rangle,\label{k-n}
\end{equation}
\begin{equation}
K_{0}|k,n\rangle=(k+n)|k,n\rangle.\label{k0n}
\end{equation}

In analogy to the harmonic oscillator coherent states, Perelomov defined the standard $SU(1,1)$
coherent states as \cite{Perellibro}
\begin{equation}
|\zeta\rangle=D(\xi)|k,0\rangle=(1-|\zeta|^2)^k\sum_{s=0}^\infty\sqrt{\frac{\Gamma(n+2k)}{s!\Gamma(2k)}}\zeta^s|k,s\rangle,\label{PCN}
\end{equation}
where $|k,0\rangle$ is the lowest normalized state. In this expression, $D(\xi)$ is displacement operator for this group defined as
$D(\xi)=\exp(\xi K_{+}-\xi^{*}K_{-})$, where $\xi=-\frac{1}{2}\tau e^{-i\varphi}$, $-\infty<\tau<\infty$ and $0\leq\varphi\leq2\pi$. The so-called normal form of the displacement operator is given by \begin{equation}
D(\xi)=\exp(\zeta K_{+})\exp(\eta K_{0})\exp(-\zeta^*K_{-})\label{normal},
\end{equation}
where  $\zeta=-\tanh(\frac{1}{2}\tau)e^{-i\varphi}$ and $\eta=-2\ln \cosh|\xi|=\ln(1-|\zeta|^2)$ \cite{Gerry}. This expression
is the analogue of equation (\ref{disentangled}) for the Weyl operator.

The $SU(1,1)$ Perelomov number coherent states
were introduced by Gerry and are defined by the following expression  \cite{gerryberry}
\begin{equation}
|\zeta,k,n\rangle=D(\xi)|k,n\rangle=\exp(\zeta K_{+})\exp(\eta
K_{3})\exp(-\zeta^* K_{-})|k,n\rangle\label{defPCNS}.
\end{equation}
By using the BCH formula of equation (\ref{BCH}), it has been shown that the similarity transformation of the operators $K_{\pm}$ are \cite{nosotros1}
\begin{equation}
L_{+}=D(\xi)K_{+}D^{\dag}(\xi)=-\frac{\xi^{*}}{|\xi|}\alpha
K_{0}+\beta\left(K_{+}+\frac{\xi^{*}}{\xi}K_{-}\right)+K_{+},
\end{equation}
\begin{equation}
L_{-}=D(\xi)K_{-}D^{\dag}(\xi)=-\frac{\xi}{|\xi|}\alpha
K_{0}+\beta\left(K_{-}+\frac{\xi}{\xi^{*}}K_{+}\right)+K_{-},
\end{equation}
These action of these on the $SU(1,1)$ Perelomov number coherent states is
\begin{equation}
L_{+}|\zeta,k,n\rangle=\sqrt{(n+1)(2k+n)}|\zeta,k,n+1\rangle,\quad\quad
L_{-}|\zeta,k,n\rangle=\sqrt{n(2k+n-1)}|\zeta,k,n-1\rangle.\label{ln}
\end{equation}
Thus, $L_{\pm}$ act as ladder operators for these number coherent states. Also, the most general form of theses states
on the Fock space was calculated as follows \cite{nosotros1}
\begin{eqnarray}
|\zeta,k,n\rangle &=&\sum_{s=0}^\infty\frac{\zeta^s}{s!}\sum_{j=0}^n\frac{(-\zeta^*)^j}{j!}e^{\eta(k+n-j)}
\frac{\sqrt{\Gamma(2k+n)\Gamma(2k+n-j+s)}}{\Gamma(2k+n-j)}\nonumber\\
&&\times\frac{\sqrt{\Gamma(n+1)\Gamma(n-j+s+1)}}{\Gamma(n-j+1)}|k,n-j+s\rangle.\label{PNCS}
\end{eqnarray}
These states are the analogue for the $SU(1,1)$ group to those given by Nieto for the harmonic oscillator in equation (\ref{nieto}).

\section{A charged particle in a uniform magnetic field in the Landau gauge.}

The stationary Schr\"odinger equation of a charged particle in a uniform magnetic field $\vec{B}$ is given by
\begin{equation}
H\Psi=\frac{1}{2\mu}\left(p+\frac{e}{c}\vec{A}\right)^2\Psi=E\Psi,\label{Hamil}
\end{equation}
where $\vec{A}$ is the vectorial potential, related to the magnetic field as $\vec{B}=\nabla\times \vec{A}$.
This vector potential does not describe the magnetic field in a unique way, since the magnetic field remains
invariant against gauge transformations $\vec{A}\rightarrow \vec{A}'=\vec{A}+\nabla g$,
where $g$ is a time independent scalar field \cite{Capri}. We can choose the vector potential as $\vec{A}=\frac{1}{2}\vec{B}\times\vec{r}$
and our coordinate system so that the z-axis is parallel to $\vec{B}$. Then,
\begin{equation}
\vec{A}=-\frac{B}{2}(y,-x,0).\label{symmetric}
\end{equation}
This choice is known as the symmetric gauge. If we make the gauge transformation with $g=-\frac{B}{2}xy$,
we obtain the following vector potential
\begin{equation}
\vec{A}'=-B(y,0,0).
\end{equation}
This choice of the vector potential is known as the Landau gauge. With this gauge the equation (\ref{Hamil}) becomes \cite{Capri}
\begin{equation}
-\frac{\hbar^2}{2m}\frac{d^2\psi}{dy^2}+\frac{m\omega^2}{2}(y-d)^2\psi=\left(E-\frac{\hbar^2k_z^2}{2m}\right)\psi,\label{haml}
\end{equation}
where the Larmor frequency $\omega$ and $d$ are defined as
\begin{equation}
\omega=\frac{eB}{mc},\quad\quad d=\frac{\hbar ck_x}{eB}.
\end{equation}
By introducing the harmonic oscillator operators $a, a^{\dag}$
\begin{equation}
a=\sqrt{\frac{m\omega}{\hbar}}y+\frac{i}{\sqrt{2m\omega\hbar}}p_y, \quad\quad a^{\dag}=\sqrt{\frac{m\omega}{\hbar}}y-\frac{i}{\sqrt{2m\omega\hbar}}p_y,
\end{equation}
we can write the equation (\ref{haml}) as follows
\begin{equation}
H\psi=\mu\left(a^{\dag}a+\frac{1}{2}\right)\psi+\nu\left(a+a^{\dag}\right)\psi=\left(E-\frac{\hbar^2k_z^2}{2m}-\frac{m\omega^2}{2}d^2\right)\psi,
\end{equation}
where
\begin{equation}
\mu=\hbar\omega,\quad\quad\quad \nu=-\sqrt{\frac{\hbar m\omega^3}{2}}d.
\end{equation}
If we make the definition $\epsilon=E-\frac{\hbar^2k_z^2}{2m}-\frac{m\omega^2}{2}d^2$ and in order to diagonalize this Hamiltonian, we apply the tilting transformation with the displacement operator as follows \cite{nosotros1,nosotros2}
\begin{equation}
D^{\dag}(\alpha)HD(\alpha)D^{\dag}(\alpha)\psi=\epsilon D^{\dag}(\alpha)\psi.
\end{equation}
From equations (\ref{A}) the tilted Hamiltonian $H'=D^{\dag}(\alpha)HD(\alpha)$ becomes
\begin{equation}
H'=\mu\left(a^{\dag}a+|\alpha|^2+\frac{1}{2}\right)-\nu(\alpha+\alpha^*)+a^{\dag}(\alpha\mu-\nu)+a(\alpha^*\mu-\nu).
\end{equation}
If we choose the coherent state parameters $y_0=d$ and  $p_{y_0}=0$ we obtain that the tilted Hamiltonian reduces, up to a constant factor, to that
of the one-dimensional harmonic oscillator
\begin{equation}
H'=\mu\left(a^{\dag}a+|\alpha|^2+\frac{1}{2}\right)-\nu(\alpha+\alpha^*).
\end{equation}
Thus, the energy spectrum for a charged particle in an uniform magnetic field is
\begin{equation}
E=\left(n+\frac{1}{2}\right)\hbar\omega+\frac{\hbar^2k_z^2}{2m}.\label{spectrum}
\end{equation}
The wave function $\psi$ is obtained by applying the displacement operator $D(\alpha)$ to the
harmonic oscillator wave functions $\psi'$. Thus, from equation (\ref{dfunction})
\begin{equation}
\psi=D(\alpha)\psi'=N_1e^{-\frac{(y-y_0)^2}{2\lambda^2}}H_n\left(\frac{y-y_0}{\lambda}\right).\label{eigen}
\end{equation}
In this expression $N_1$ is a normalization constant and $\lambda$ is the magnetic length $\lambda=\left(\frac{\hbar c}{eB}\right)^{1/2}$. The eigenfunction for the general problem $\Psi$ is obtained by adding the free particle term $e^{i(k_xx+k_zz)}$ to equation (\ref{eigen}). Therefore, we have showed that the eigenfunctions of a charged particle in an uniform magnetic field are the harmonic oscillator number coherent states. The treatment developed in this section can be also applied to the problem of a charged particle in a pure electric field or in a magnetic and electric field. With a proper choice of the
coherent states parameters it can be shown that the harmonic oscillator number coherent states are the eigenfunctions of these problems.

\section{A charged particle in a uniform magnetic field in the symmetric gauge and its $SU(1,1)$ number coherent states}

In the symmetric gauge (equation (\ref{symmetric})) the Schr\"odinger equation of a charged particle in a uniform magnetic field is
\begin{equation}
H\psi=\frac{-\hbar^2}{2\mu}\nabla^2\psi+\frac{eB}{2\mu c}L_z\psi+\frac{e^2B^2}{8\mu c^2}(x^2+y^2)\psi=E\psi.
\end{equation}
If we consider the wave function $\psi(r)=U(\rho)e^{im\phi}e^{ikz}$ $(m=0,1,2,...)$ in cylindrical coordinates, the Schr\"odinger
equation for a charged particle in an external magnetic field remains
\begin{equation}
\left[\frac{d^2}{d\rho^2}+\frac{1}{\rho}\frac{d}{d\rho}-\frac{m^2}{\rho^2}-\frac{e^2B^2}{4\hbar^2c^2}\rho^2
+\frac{2mE}{\hbar^2}-\frac{eBm}{\hbar c}-k^2\right]U(\rho)=0.\label{schrolandau}
\end{equation}
By performing the change of variable $x=\sqrt{\frac{eB}{2\hbar c}}\rho$ in the above equation we obtain
\begin{equation}
\left[\frac{d^2}{dx^2}+\frac{1}{x}\frac{d}{dx}-\frac{m^2}{x^2}+(\lambda-x^2)\right]U(x)=0,
\end{equation}
where we have introduced the variable $\lambda$ defined as
\begin{equation}
\lambda=\frac{4\mu c}{eB\hbar}\left(E-\frac{\hbar^2k^2}{2\mu}\right)-2m.
\end{equation}
The $su(1,1)$ Lie algebra for this problem is well known and the generators for its realization are given by \cite{Gur}
\begin{equation}
K_{\pm}=\frac{1}{2}\left(\pm x\frac{d}{dx}-x^2+2K_0\pm 1\right),
\end{equation}
\begin{equation}
K_0=\frac{1}{4}\left(-\frac{d^2}{dx^2}-\frac{1}{x}\frac{d}{dx}+\frac{m^2}{x^2}+x^2\right).
\end{equation}
Moreover, by defining $y=x^2$, the normalized wave functions are
\begin{equation}
U_n(y)=\sqrt{\frac{2n!}{(n+m)!}}e^{-y/2}y^{m/2}L_n^{m}(y),\label{wavelandau}
\end{equation}
which are the Sturmian functions for the unitary irreducible representations of the $su(1,1)$ Lie algebra.
Also, the Bargmann index $k$ is $k=m/2+1/2$, and the other group number is just the radial quantum number $n$.
Therefore we can construct the $SU(1,1)$ Perelomov number coherent states for this problem by substituting equation (\ref{wavelandau})
into equation (\ref{PNCS}). By interchanging the order of summations and using the properties $48.7.6$ and $48.7.8$ of reference
\cite{hansen} we obtain
\begin{eqnarray}
\psi_{n,m}&=&\sqrt{\frac{2\Gamma(n+1)}{\Gamma(n+m+1)}}\frac{(-1)^n}{\sqrt{\pi}}e^{im\phi}
\frac{(-\zeta^*)^n(1-|\zeta|^2)^{\frac{m}{2}+\frac{1}{2}}(1+\sigma)^n}{(1-\zeta)^{m+1}}\nonumber\\
&&\times e^{-\frac{\rho^2(\zeta+1)}{2(1-\zeta)}}\rho^{m}L_n^{m}\left(\frac{\rho^2\sigma}{(1-\zeta)(1-\sigma)}\right),\label{PNCSL}
\end{eqnarray}
where we have defined
\begin{equation}
\sigma=\frac{1-|\zeta|^2}{(1-\zeta)(-\zeta^*)}.
\end{equation}
These are the $SU(1,1)$ Perelomov number coherent states of a charged particle in a magnetic field. As a particular case of this result we can see that for $n=0$ these states reduces to the standard Perelomov coherent states, presented in reference \cite{apolonica}. These states are significant in quantum optics, since a particular case of them are the eigenfunctions of the non-degenerate parametric amplifier \cite{nosotros1}.

\section{$SU(1,1)$ contraction to the Heisenberg-Weyl group.}
In this section, we will contract the $su(1,1)$ Lie algebra to the $h(4)$ algebra of the harmonic oscillator. The proceeding developed here is
analogue to that presented by Arecchi in reference \cite{arecchi} for the $su(2)$ algebra. Thus, we define the following transformation
\begin{equation}
\begin{pmatrix}
h_+\\h_-\\h_0\\h_I
\end{pmatrix}=
\begin{pmatrix}
c & 0 & 0 & 0\\0 & c & 0 & 0\\0 & 0 & 1 & -\frac{1}{2c^2}\\0 & 0 & 0 & 1
\end{pmatrix}
\begin{pmatrix}
K_+\\K_-\\K_0\\K_I
\end{pmatrix}.
\end{equation}
These new operators $h$ satisfy the following commutation relationships
\begin{equation}
[h_0,h_{\pm}]=\pm cK_{\pm},\quad\quad [h_-,h_+]=2c^2K_0,\quad\quad [\vec{h},h_I]=0.
\end{equation}
In the limit $c\rightarrow0$ this transformation becomes singular. However, the commutation relationships
are well defined and become
\begin{equation}
[h_0,h_{\pm}]=\pm h_{\pm},\quad\quad [h_-,h_+]=h_0,\quad\quad [\vec{h},h_I]=0,
\end{equation}
which is nothing but the $h(4)$ algebra with the definition
\begin{equation}
\lim_{c\rightarrow0}h_0=n=a^{\dag}a, \quad\quad \lim_{c\rightarrow0}h_+=a^{\dag}, \quad\quad\lim_{c\rightarrow0}h_-=a.
\end{equation}
Also, in order to contract the displacement operator $D(\xi)$ to the Weyl operator $D(\alpha)$ the
coherent state parameters must satisfy
\begin{equation}
\lim_{c\rightarrow0}\frac{\xi}{c}=\alpha, \quad\quad  \lim_{c\rightarrow0}\frac{\xi^*}{c}=\alpha^*.
\end{equation}
To obtain the relationship between the contraction parameter $c$ and the group number $k$ we apply the
$h_0$ operator to an arbitrary $su(1,1)$ state $|n,k\rangle$
\begin{equation}
h_0|n,k\rangle=\left(K_0-\frac{1}{2c^2}K_I\right)|n,k\rangle=\left(n+k-\frac{1}{2c^2}\right)|n,k\rangle.
\end{equation}
If we demand that this eigenvalue must vanish for the lowest state $|0,k\rangle$ we obtain
\begin{equation}
\lim_{c\rightarrow0}\left(k-\frac{1}{2c^2}\right)=0.
\end{equation}
Thus, in the limit $c\rightarrow0$, $c=\sqrt{\frac{1}{2k}}$ and the $su(1,1)$ irreducible unitary representations
contract to the $h(4)$ irreducible unitary representations. The relationship between the states of both groups can be obtained by defining the state
\begin{equation}
|\infty,n\rangle=\lim_{c\rightarrow0}|n,k\rangle.
\end{equation}
With this definition we obtain
\begin{eqnarray}
a^{\dag}a|\infty,n\rangle=\lim_{c\rightarrow0}\left(K_0-\frac{1}{2c^2}\right)|n,k\rangle
=\lim_{c\rightarrow0}\left(n+k-\frac{1}{2c^2}\right)|n,k\rangle
=n|\infty,n\rangle.
\end{eqnarray}
In a similar way we obtain
\begin{equation}
a^{\dag}|\infty,n\rangle=\sqrt{n+1}|\infty,n+1\rangle\quad\quad\quad a|\infty,n\rangle=\sqrt{n}|\infty,n-1\rangle.
\end{equation}
Therefore, the Perelomov number coherent states contract to the harmonic oscillator number coherent states, since
\begin{equation}
|\alpha\rangle=\lim_{c\rightarrow0}|\zeta,n,k\rangle=\lim_{c\rightarrow0}\left(1-|\zeta|^2\right)^k e^{\xi}K_+|n,k\rangle\\
=\lim_{c\rightarrow0}\left(1-c^2\alpha\alpha^*\right)^{1/2c^2}e^{\alpha a^{\dag}}|0\rangle\\
=e^{-|\alpha|^2}e^{\alpha a^{\dag}}|0\rangle.
\end{equation}
In our problem, this implies that the $SU(1,1)$ Perelomov number coherent states of a charged particle in a magnetic field of equation (\ref{PNCSL}), under the contraction of the $SU(1,1)$ group, reduce to the number coherent states of the harmonic oscillator of equation (\ref{eigen}). In reference \cite{blasone}, the authors studied the contraction of the $SU(1,1)$ group to the quantum harmonic oscillator. Moreover, they shown that one advantage of
working with $SU(1,1)$ is that its representation Hilbert space is infinite-dimensional, thus it does not change dimension in the contraction limit, as it happens for the $SU(2)$ case.

\section{Concluding remarks}

We applied the generalized number coherent states theory to study the problem of a charged particle in the Landau and symmetric gauge. We showed that for the Landau gauge, the eigenfunctions for the Landau level states can be represented in terms of the harmonic oscillator coherent states. For the symmetric gauge we study the eigenfunctions of this problem in cylindrical coordinates and we constructed the $SU(1,1)$ Perelomov number coherent states in a closed way. We show that under a contraction of the $SU(1,1)$ group, the Perelomov number coherent states are reduced to the number coherent states of the harmonic oscillator, related to the Heisenberg-Weyl group.

It is important to note that the tilting transformation method used in this work has been applied to more novel problems, as the non-degenerate parametric amplifier \cite{nosotros1}, the problem of two coupled oscillators \cite{nosotros2}, and the generalized MICZ-Kepler problem \cite{nosotros3}.

\section*{Acknowledgments}
This work was partially supported by SNI-M\'exico, COFAA-IPN,
EDI-IPN, EDD-IPN, SIP-IPN project number $20170632$.

\end{document}